\documentclass[aps,pra,showpacs,twocolumn]{revtex4-1}
\usepackage{amsmath,amsfonts,amssymb}
\usepackage{graphicx,float,calc}
\usepackage{color,bm}
\usepackage{ulem}
\usepackage[colorlinks,urlcolor=blue,citecolor=blue,linkcolor=blue]{hyperref}

\def\dd{\mathrm{d}}
\def\ii{\mathrm{i}}
\def\ee{\mathrm{e}}

\begin{document}

\author{Zhen Zheng}
\thanks{zhenzhen.dr@hku.hk}
\author{Z. D. Wang}
\thanks{zwang@hku.hk}

\affiliation{Department of Physics and Center of Theoretical and Computational Physics, The University of Hong Kong, Pokfulam Road, Hong Kong, China}
\title{Manipulating Cooper pairs with a controllable momentum in periodically driven degenerate Fermi gases}

\begin{abstract}

We here present an experimentally feasible proposal for manipulating Cooper pairs in degenerate Fermi gases trapped by an optical lattice.
Upon introducing an \textit{in situ} periodically driven field,
the system may be described by an effective time-independent Hamiltonian,
in which the Cooper pairs, generated by the bound molecule state in Feshbach resonance, host a nonzero center-of-mass momentum.
The system thus processes a crossover from a Bardeen-Cooper-Schrieffer (BCS) superfluid phase to a Fulde-Ferrell (FF) one.
Furthermore, the magnitude and direction of the Cooper pairs in the synthetic FF superfluids
are both directly controllable via the periodically driven field.
Our proposal offers a reliable and feasible scenario for manipulating the Cooper pairs in cold atoms,
serving as a tunable as well as powerful platform for quantum-emulating and exploring the FF superfluid phase.

\end{abstract}


\maketitle

\section{Introduction}

Manipulation of cold atoms via optical techniques
has intensively been  studied \cite{quantum-simulation-rev},
which offers a powerful experimental tool for synthesizing
many interesting phases and models that are hardly accessible in conventional solid systems.
On one hand,  Raman transitions can be implemented to couple pseudo-spins of cold atoms,
and a variety of experiments have been performed for artificial Abelian and non-Abelian gauge fields
\cite{raman-rev-2011,raman-rev-2014}.
On the other hand, the periodically driven cold-atom system
opens an alternative window to achieve 
modulated couplings \cite{floquet-tunnel,floquet-soc-2012,floquet-soc-2014} or interactions \cite{floquet-interact-2012,floquet-interact-2016}.
This technique 
leads to a so-called 
Floquet engineering \cite{floquet-rev-2015,floquet-rev-2017},
and has been employed in quantum simulations,
such as the Mott-insulator to superfluid transition \cite{mott-sf},
topological insulator \cite{lindner2011floquet,floquet-topo-ins,qizhou-shaken},
ferromagnetic transition \cite{parker2013direct},
artificial magnetic fields \cite{floquet-gauge},
and superlfluid Ising transition \cite{shaken-ising}.

Generally, the Floquet engineering is based on enforcing periodically time-dependent 
external fields or mechanical deformations on the original static system.
By applying an \textit{in situ} perturbative driven field,
the system may be captured by an effective Hamiltonian
with various modulated parameters \cite{floquet-eff-hami-1988,floquet-eff-hami-2003,floquet-eff-hami-2010,floquet-eff-hami-2015}.
This invokes an idea of controlling and manipulating Cooper pairs in cold atoms
with these tunable parameters.
In particular, we are motivated to search for a reliable and feasible experimental scheme for
Floquet-engineering a superfluid phase with the fully controllable pairing momentum, 
which is known as the Fulde-Ferrell (FF) phase \cite{fflo-1st-ff,fflo-1st-lo}.

This paper is organized as follows.
In Sec.\ref{sec-model}, we start with a two-channel model Hamiltonian in Feshbach resonance,
and address our motivation of this paper for manipulation of Cooper pairs as well as
the synthetic FF superfluids.
In Sec.\ref{sec-driven},
we elaborate that the introduction of the periodically driven field will result in a artificial controllable pairing momentum, which
yields the FF superfluid phase.  
Its existence as the ground state of the lattice system will be shown in Sec.\ref{sec-numeric}.
 We detail how to realize the periodically driven field via current experimental techniques in Sec.\ref{sec-exp}.
The features of the synthetic FF superfluids in our proposal are discussed in Sec.\ref{sec-diss}.
We make a brief summary the in the last Sec.\ref{sec-summary}.

\section{Model Hamiltonian}\label{sec-model}

We consider a degenerate Fermi gas with two pseudo-spins trapped in a three dimensional (3D) optical lattice.
In real experiments, the interaction between two atoms with opposite spins is realized via Feshbach resonance,
in which the two atoms collide and bound to a bosonic molecule state.
This system may be described by a two channel model  Hamiltonian 
composed of three parts \cite{lattice-bcs-bec-crossover},
\begin{equation}
H = H_c + H_b + H_{bc} ~.\label{eq-h-real}
\end{equation}
The first part $H_c$ describes two free atoms in the open channel of the scattering process,
\begin{equation}
H_c = \int \dd \bm{r} \sum_{\sigma=\uparrow\downarrow}
\psi_\sigma^\dag(x) [-\bm{\nabla}^2/2m+V_L(\bm{r}) - \mu]\psi_\sigma(x) ~,
\end{equation}
where $\psi_{\sigma}^\dag,\psi_{\sigma}$ are the creation and annihilation operators of the fermionic atoms. 
$V_L(\bm{r})=V_L\sin^2(k_Lx)+V_L\sin^2(k_Ly)+V_L\sin^2(k_Lz)$ is the lattice trap potential.
$k_L=\pi/a$ with $a$ as the lattice constant.
In the whole paper we assume $\hbar=1$.
$\mu$ is the chemistry potential of fermions.
The second part $H_b$ is the molecule state of the close channel,
\begin{equation}
H_b = \int \dd \bm{r} \,
\varphi^\dag(x) [-\bm{\nabla}^2/2m+V_L(\bm{r}) - 2\mu]\varphi(x) ~,
\end{equation}
where $\varphi^\dag,\varphi$ are the creation and annihilation operators of the bosonic molecule state.
$2\mu$ is introduced due to the number conservation.
The last part $H_{bc}$ corresponds to the coupling of the two channels.
For simplicity, we here consider the contact interaction with strength $g$, so that
$H_{bc}$ is expressed as
\begin{equation}
H_{bc} = g\int \dd \bm{r} \, \varphi^\dag(\bm{r}) \psi_\uparrow(\bm{r}) \psi_\downarrow(\bm{r}) ~.
\end{equation}

In the lattice system, we can study the Hamiltonian using the tight-binding approximation.
For details, we expand $\psi_\sigma$ and $\varphi$ by Wannier wave functions $W(\bm{r})$,
\begin{equation}
\psi_\sigma(\bm{r}) = \sum_j W(\bm{r}-\bm{r}_j) c_{j\sigma} \,,\,
\varphi(\bm{r}) = \sum_j W(\bm{r}-\bm{r}_j) b_{j} \,.
\end{equation}
Here we denote $c$ and $b$ as operators of the fermion and molecule state.
Then Hamiltonian (\ref{eq-h-real}) is represented as
\begin{equation}
\hat{H} = \hat{H}_0+ \hat{H}_1~,
\label{eq-h-site}
\end{equation}
where $\hat{H}_0$ is the intra-channel Hamiltonian
\begin{equation}
\hat{H}_0 = \sum_{j,\sigma} (E_b-2\mu)b_j^\dag b_j - \mu c_{j\sigma}^\dag c_{j\sigma} 
-( t c_{j\sigma}^\dag c_{j+1\sigma} + H.c.)
\label{eq-h-site-0}
\end{equation}
and $\hat{H}_1$ describes the inter-channel coupling
\begin{equation}
\hat{H}_1= \sum_{j} U b_j^\dag c_{j\uparrow} c_{j\downarrow} + H.c.
\label{eq-h-site-1}
\end{equation}
$H.c.$ is the Hermitian conjugate.
$t$ is the tunneling magnitude stemmed from the kinetic energy in $H_c$.
$E_b$ is the bound energy and can be shifted via magnetic fields,
which plays the key role for a controllable atomic interaction via Feshbach resonance.
The interaction strength $U$ is given by
$U= g\int \dd\bm{r}\, W^*(\bm{r})[W(\bm{r})]^2$.

In this paper, we take the mean field method to study the lattice system,
since it can give a clear physics picture and capture qualitative features of the 3D system.
In an ordinary picture, 
we can replace the molecular field by its mean value $b_j\approx\langle b_j \rangle = b$ in Eq.(\ref{eq-h-site-1}).
It reveals the $s$-wave Cooper pairs is dominant whose presence characterizes the Bardeen-Cooper-Schrieffer (BCS) superfluid phase.
This can be shown by Fourier transforming the Hamiltonian (\ref{eq-h-site-1}) into the momentum space,
\begin{align}
\tilde{H}_1 = \sum_{\bm{k}} U b^* c_{\bm{k}\uparrow} c_{-\bm{k}\downarrow}+ H.c.
\label{eq-h-momentum}
\end{align}
The reason of using such a mean-field solution is because, for extremely cold atoms, the bosonic molecule state will condensate on the state with zero momentum.
A recent study shows that the molecule state can acquire a nonzero center-of-mass momentum via the Raman transition to auxiliary levels \cite{dark-state-fflo}.
The Cooper pairs in that system thus host a nonzero pairing momentum,
yielding the realization of the FF superfluid phase.
It inspires us with an interesting question:
is there a simpler proposal for realizing the FF superluids without manipulating the molecule state
by optical methods?
On the other hand, recent investigations on Floquet engineering 
have addressed how to modulate single-particle fields with time-dependent fields.
This motivates us to search for an alternative proposal with a periodically driven system.

\section{Periodically driven engineering}\label{sec-driven}

We consider a perturbative locally and periodically driven field as follows,
\begin{equation}
V(t) = \frac{\Delta}{2} \cos(\omega t + \phi_j) -\frac{\omega}{2} ~. \label{eq-dr-potential}
\end{equation}
$\Delta$ and $\omega$ are the magnitude and frequency of the periodically driven field.
The phase $\phi_j=j\eta \pi$ depends on the site index in accompany with a controllable parameter $\eta$.
For simplicity, in the whole paper, we consider its projection along the $x$ direction $\phi_j=\phi_{j_x}=j_x\eta \pi$ instead ($j_x$ is the site index along the $x$ direction).
Adding $V(t)$ to the lattice system, its Hamiltonian is given by
\begin{equation}
\mathcal{H} = \hat{H}_0+\hat{H}_1+H_t ~,\quad
H_t =\sum_{j,\sigma} V(t) c_{j\sigma}^\dag c_{j\sigma} ~.
\label{eq-cal-h-site}
\end{equation}
Here $\hat{H}_{0,1}$ have been given in Eqs.(\ref{eq-h-site-0})-(\ref{eq-h-site-1}).
In order to obtain a time-independent effective Hamiltonian,
we make the following unitary transformation to eliminate the the periodically driven term $H_t$:
\begin{equation}
\mathcal{U} = \exp\Big[\frac{\ii}{2}\sum_{j,\sigma} \Omega_j(t) c_{j,\sigma}^\dag c_{j,\sigma} \Big]
\end{equation}
with $\Omega_j(t) = \frac{\Delta}{\omega}\sin(\omega t+\phi_j) -\omega t$.
In the rotating frame, the effective Hamiltonian is given by
\begin{equation}
\mathcal{H}' = \mathcal{U} \mathcal{H} \mathcal{U}^\dag - \ii \mathcal{U} \partial_t \mathcal{U}^\dag
= \hat{H}_0' + \hat{H}_1'
~.\label{eq-cal-h-site-2nd}
\end{equation} 
where
\begin{align}
&\hat{H}_0' = \mathcal{E}_0-t\sum_{j,\sigma} \ee^{\ii [\Omega_j(t)-\Omega_{j+1}(t)]/2} c_{j\sigma}^\dag c_{j+1\sigma} + H.c. \\
&\hat{H}_1' = U\sum_{j} \ee^{-\ii \Omega_j(t)} b_j^\dag c_{j\uparrow} c_{j\downarrow} + H.c.
\end{align}
and $\mathcal{E}_0=\sum_{j,\sigma} (E_b-2\mu)b_j^\dag b_j - \mu c_{j\sigma}^\dag c_{j\sigma}$.
Using Bessel expansion $\ee^{\ii z\sin\theta}=\sum_n J_n(z) \ee^{\ii n\theta}$ ($J_n$ denotes the $n$-th order Bessel function), we can get
\begin{align}
&\hat{H}_0' = \mathcal{E}_0-t\sum_{j,n,n'}J_n(\Delta/\omega)J_{n'}(\Delta/\omega) \ee^{\ii (n-n')\omega t +\ii (n\phi_j-n'\phi_{j+1})} \notag\\
&\qquad \times c_{j\sigma}^\dag c_{j+1\sigma} + H.c.\\
&\hat{H}_1' = U\sum_{j,n}J_n(\Delta/\omega) \ee^{-\ii (n-1)\omega t -\ii n\phi_j} b_j^\dag c_{j\uparrow} c_{j\downarrow} + H.c.
\end{align}
In practice, if we tune $\omega\gg \Delta$, we can neglect rapidly oscillating terms,
and obtain
\begin{align}
&\hat{H}_0' \approx \mathcal{E}_0-\tilde{t} \sum_j c_{j\sigma}^\dag c_{j+1\sigma} +H.c. \\
&\hat{H}_1' \approx \tilde{U} \sum_{j} \ee^{-\ii\phi_j} b_j^\dag c_{j\uparrow} c_{j\downarrow} + H.c.
\end{align}
where we have denoted
\begin{equation}
\tilde{t}\equiv t [J_0(\Delta/\omega)]^2 ~,\qquad
\tilde{U}\equiv UJ_1(\Delta/\omega) ~.
\end{equation}
We submit $\hat{H}_{0,1}'$ into Eq.(\ref{eq-cal-h-site-2nd}) and make the mean-field approximation $b_j\approx b$.
In the momentum space, we obtain the final time-independent form of the effective Hamiltonian,
\begin{equation}
H_\mathrm{eff} = \sum_{\bm{k},\sigma} \xi_b + \xi_{k} c_{\bm{k}\sigma}^\dag c_{\bm{k}\sigma} 
+ \tilde{U}(b^* c_{\bm{k}\uparrow} c_{\bm{Q}-\bm{k}\downarrow}+ H.c.)
\label{eq-cal-h-momentum}
\end{equation}
with
\begin{equation}
\xi_b=(E_b-2\mu) |b|^2 ~,~
\xi_{k} = -\mu -2\tilde{t} \sum_{i=x,y,z}\cos(k_ia) ~.
\end{equation}
The Cooper pairs host a center-of-mass momentum $\bm{Q}=\eta k_L\hat{\bm{e}}_x$,
where $\hat{\bm{e}}_x$ denoting the lattice primitive vector along the $x$ direction.
As $\eta$ can be changed in the periodically driven field,
it indicate that the magnitude as well as the direction of the pairing momentum are both directly introduced and controllable via optical techniques.
By contrast, in a previous Floquet engineering proposal \cite{zz-shaken-fflo},
the FF phase emerges due to the orbit band inverse and hence its pairing momentum is fixed.

\section{Numeric results}\label{sec-numeric}

From the effective Hamiltonian (\ref{eq-cal-h-momentum}),
in the base $\Psi_{\bm{k}}=(c_{\bm{k}\uparrow},c_{\bm{Q}-\bm{k}\downarrow}^\dag)^T$,
we can write the Bogoliubov-de Gennes (BdG) Hamiltonian, 
\begin{equation}
H_\mathrm{BdG} (\bm{k}) =
\begin{pmatrix}
\xi_{\bm{k}} & \tilde{U}b \\
\tilde{U}b^* & -\xi_{\bm{Q}-\bm{k}}
\end{pmatrix} ~.
\end{equation}
The diagonalization of $H_\mathrm{BdG} (\bm{k})$ gives the spectrum of the quasi-particles:
$E^{\pm}_{\bm{k}}=\xi_{\bm{k}}^- \pm \sqrt{(\xi_{\bm{k}}^+)^2+ \tilde{U}^2|b|^2}$.
We denote $\xi_{\bm{k}}^\pm=(\xi_{\bm{k}}\pm \xi_{\bm{Q}-\bm{k}})/2$.
The system energy is calculated by $\mathcal{E}=\langle H_\mathrm{eff} \rangle$, which is written as
\begin{equation}
\mathcal{E} = \sum_{\bm{k},\gamma=\pm} E^{\gamma}_{\bm{k}} \Theta(-E^{\gamma}_{\bm{k}}) + \xi_{\bm{k}}+ \xi_b ~.
\end{equation}
Here $\Theta(x)$ is the Heaviside step function that describes the Fermi distribution at zero temperature.
The gap and number equations can be obtained by \cite{huhui-2006}
\begin{equation}
\frac{\partial \mathcal{E}}{\partial |b|}=0 ~,\qquad
\frac{\partial \mathcal{E}}{\partial \mu}=-n ~. \label{eq-consistent-eqs}
\end{equation}
Here $n$ is the filling factor and initial determined when preparing the degenerate Fermi gas.
The superfluid order parameter $|b|$ and the chemistry potential $\mu$
are obtained simultaneously by self-consistent solving Eq.(\ref{eq-consistent-eqs}).
When $|b|\neq0$, the ground state is the superfluid phase,
meanwhile, is the BCS(FF)-type if we set $\eta=$$(\neq)0$.
When $|b|$ vanishes, the system is a normal gas.

\begin{figure}[t]
\centering
\includegraphics[width=0.48\textwidth]{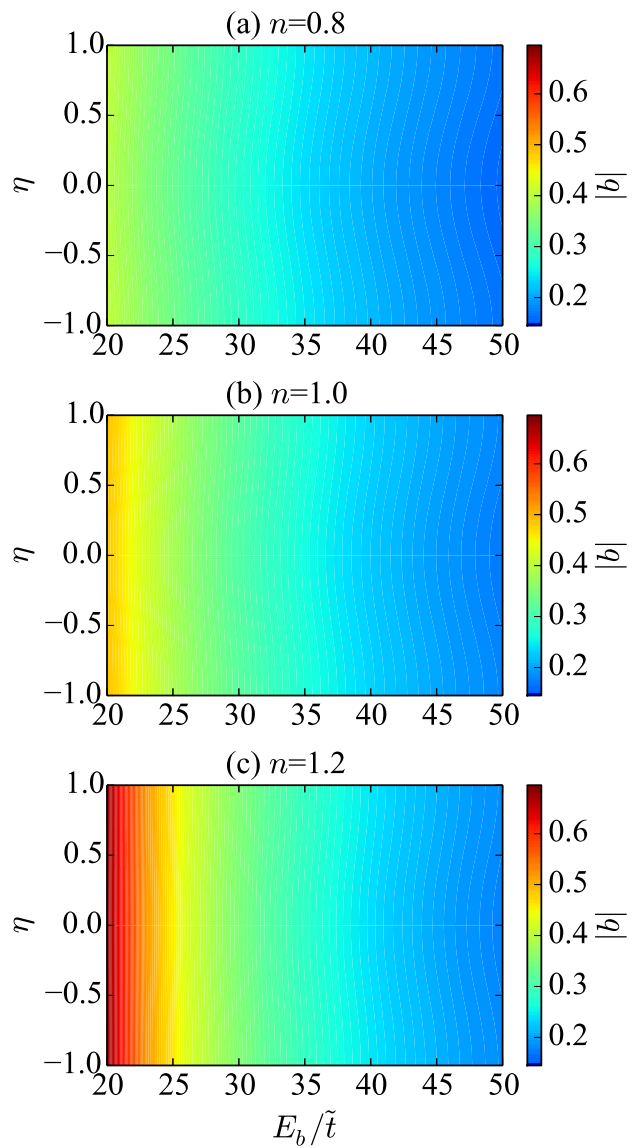}
\caption{Superfluid order parameter $|b|$ as a function in the $\eta$-$E_b$ plane at different filling factors.
The color visualizes the value of $|b|$.
We set $\tilde{U}=20\tilde{t}$ in the numeric calculations.}
\label{fig-order}
\end{figure}

In Fig. \ref{fig-order},
we plot the evolution of the superfluid order parameter $|b|$ in the $\eta$-$E_b$ plane.
Both two parameters $\eta$ and $E_b$ are experimentally tunable.
We can see the superfluid phase is still present when $\eta\neq0$,
yielding the existence of the FF superfluids as the ground state of the lattice system.
$|b|$ is dominated by $E_b$ as well as $n$, and increase monotonically with the decrease of $E_b$,
which processes a BCS--Bose-Einstein-condensate(BEC) crossover.
By contrast, it changes insensitive and slightly with $\eta$ and is independent from $\eta$'s sign.
It is straight forward to understand this phenomenon,
because the lattice system is spin degenerate despite that $\bm{Q}$ is artificially introduced.
The Fermi surfaces of opposite spins do not split nor deform,
implying the mechanics of the FF-type Cooper pair resembles the BCS-type one's.

\section{Experimental realization}\label{sec-exp}

The proposal of the controllable Cooper pairs is ready to be realized in current experimental technique.
The time-dependent external potential $V(t)$, see Eq.(\ref{eq-dr-potential}), is composed of two terms.
The first term gives the periodically driven field.
It can be introduced by add a group of counter-propagating lasers, whose wavelength is $\lambda_L/\eta$
along the $x$ direction.
Here $\lambda_L$ is the wavelength of lasers that construct the optical lattice.
The counter-propagating lasers can give rise to a perturbative time-dependent lattice potential 
$\frac{\Delta}{2}\cos(\eta k_Lx +\omega t)$, where $\omega$ satisfies $\omega=2\pi c\eta/\lambda_L$ ($c$ is the light speed).
We should note that laser strength $\Delta/2 \ll V_L$, which guarantees the driven field does not change the lattice configuration.

The second term in $V(t)$ can be engineered intuitively by the AC-Stark shift to the fermions via an additional laser.
As this term is a local static field for the atoms, 
it can be recognized as the additional energy level offset between the open and close channels in Feshbach resonance.
Therefore, the generation of the the second term in $V(t)$ is equivalent to shift the bound energy of the molecule state $E_b$ to $E_b+\omega/2$,
which can be compensated by the magnetic field used in Feshbach resonance.

\section{Discussions}\label{sec-diss}

The Cooper pairing momentum $\bm{Q}$ of the FF superfluids is originated from the external periodically driven field.
We emphasize that the synthetic FF superfluids in our proposal bear the following two features:
(i) the magnitude of $\bm{Q}$ is proportional to the parameter $\eta$, 
which stems from the wavelength of lasers that generate the driven field;
(ii) the direction of $\bm{Q}$ is governed by the same lasers' direction.
The two features differentiate this proposal from a piece of earlier Floquet engineering work \cite{floquet-fflo},
in which, like many other pieces of cold-atom works on FF superfluids,
the pairing momentum is evidenced by the self-consistently solution, 
and cannot be directly determined by the driven field.
Our proposal facilitates the manipulation of the pairing momentum in the synthetic FF phase,
providing a simpler method to directly control not only its magnitude but also the direction.

Previous cold-atom works use the Zeeman field to break the spin degeneracy,
resulting in split the Fermi surfaces of opposite spins.
In this way, a nonzero pairing momentum, i.e. the FF superfluids, is acquired.
However, it has been known that the Zeeman field suppress the superfluid order parameter, leading to a narrow region in the phase diagram \cite{huhui-2006}.
In our proposal, the engineered FF phase does not require the Zeeman field.
The lattice system is spin balanced.
The absence of Zeeman fields in our proposal will make the FF superfluids more robust against fluctuations.
It makes our proposal a promising candidate to quantum simulate and study the FF superfluids.

 It was seen in Sec.\ref{sec-driven} that
introduction of the driven field does not change the form of onsite and tunneling Hamiltonians,
except for the modulated magnitude characterized by Bessel functions.
This works even in the presence of the spin-orbital coupling or Zeeman field,
because the periodically driven field is spin independent.
Therefore in a Rashba spin-orbital coupled Fermi gas, it is possible, in absence of the in-plane Zeeman field,
to engineer topological nontrivial FF superfluids that support Majorana fermion states.
This is very different from the picture reported in previous cold-atom works \cite{qu2013topological,zhang2013topological},
in which the in-plane Zeeman field is necessary for emergence of the FF phase.

\section{Summary}\label{sec-summary}

In summary, we have proposed how to manipulate Cooper pairs in a periodically driven degenerate Fermi gas.
Different from the conventional picture, the nonzero Cooper pairing momentum is artificially introduced by optical techniques.
Its magnitude and direction are both directly designed by the driven field,
which makes our proposal more reliable and feasible for manipulating Cooper pairs.
 Since the breaking of the spin degeneracy is not required, 
the synthetic FF superfluid phase is more robust
in comparison with the previous proposals based on spin polarized gases.

\section{Acknowledgements}

This work was supported by the GRF (No.: HKU173057/17P) and CRF (No.: C6005-17G) of Hong Kong.
Z.Z. acknowledges the National Natural Science Foundation of China (Grant No. 11704367).

\bibliography{ref}

\end{document}